\documentclass[prd,aps, tightenlines, preprint, preprintnumbers, showpacs, nofootinbib,superscriptaddress]{revtex4}

\usepackage{graphicx}
\usepackage{amsmath,amsthm,amssymb}
\usepackage{mathtools}
\usepackage[usenames,dvipsnames]{color}

\begin{document}
\title{Transverse Momentum Dependent Parton Distributions at Small-x}

\author{Bo-Wen Xiao}
\affiliation{Key Laboratory of Quark and Lepton Physics (MOE) and Institute
of Particle Physics, Central China Normal University, Wuhan 430079, China}

\author{Feng Yuan}
\affiliation{Nuclear Science Division, Lawrence Berkeley National
Laboratory, Berkeley, CA 94720, USA}

\author{Jian Zhou}
\affiliation{\normalsize\it  Key Laboratory of Particle Physics and Particle Irradiation (MOE) and
School of physics, Shandong University, Jinan 250100, China}

\begin{abstract}
 We study the transverse momentum dependent (TMD) parton distributions at small-$x$ in a consistent
framework that takes into account the TMD evolution and small-$x$ evolution simultaneously.
The small-$x$ evolution effects are included by computing the TMDs at  appropriate scales in
terms of the dipole scattering amplitudes, which obey the relevant Balitsky-Kovchegov equation.
Meanwhile, the TMD evolution is obtained by resumming the Collins-Soper type large logarithms
emerged from the calculations in small-$x$ formalism into Sudakov factors.
\end{abstract} 

 \maketitle

\section{Introduction}

Transverse momentum dependent (TMD) parton distributions are among the most 
important and interesting topics to be fully investigated at the current and future 
facilities, including JLab 12 GeV upgrade, RHIC, and the planed electron-ion collider (EIC),
and have been subjects of intense studies from both theory and experiment sides
in the last decade or so. Most recent developments focus on the particular
kinematics where common interests have attracted the attentions from both
the hadron physics and heavy ion physics communities, i.e., the TMDs at small-$x$.
From the theoretical point of view, it has been shown that the TMDs at small-$x$ are unified with
the un-integrated gluon distributions (UGDs), which are widely applied
in heavy ion physics, in particular, as an important ingredient to
describe the initial conditions for heavy ion collisions at high energies.
In the last few years, there have been tremendous progresses
in connecting TMDs and small-$x$ saturation physics.

There are two different unintegrated gluon
distributions\cite{McLerran:1993ni, Kovchegov:1998bi, Collins:1981uw, Ji:2005nu, 
Kharzeev:2003wz, Bomhof:2006dp, Dominguez:2010xd}. The first gluon distribution, which is known as the
Weizs\"{a}cker-Williams (WW) gluon distribution, is calculated
from the correlator of two classical gluon fields of relativistic
hadrons~\cite{McLerran:1993ni, Kovchegov:1998bi}. The WW gluon
distribution can be defined following the conventional gluon distribution~\cite{Collins:1981uw, Ji:2005nu}
\begin{eqnarray}
xG^{(1)}(x,k_{\perp })&=&\int \frac{d\xi ^{-}d^2\xi _{\perp }}{(2\pi
)^{3}P^{+}}e^{ixP^{+}\xi ^{-}-ik_{\perp }\cdot \xi _{\perp }}   \langle P|F^{+i}(\xi ^{-},\xi _{\perp })\mathcal{L}_{\xi }^{\dagger
}\mathcal{L}_{0}F^{+i}(0)|P\rangle \ ,  \label{g1}
\end{eqnarray}%
where $F^{\mu \nu }$ is the gauge field strength tensor $F_a^{\mu \nu}$ and
$\mathcal{L}_{\xi }=\mathcal{P}\exp\{-ig\int_{\xi ^{-}}^{\infty }d\zeta
^{-}A^{+}(\zeta ,\xi _{\perp })\}$
is the gauge link in the adjoint representation.
This gluon distribution can also be defined
in the fundamental representation~\cite{Bomhof:2006dp},
\begin{eqnarray}
xG^{(1)}(x,k_\perp)&=&2\int \frac{d\xi ^{-}d\xi _{\perp }}{(2\pi )^{3}P^{+}}%
e^{ixP^{+}\xi ^{-}-ik_{\perp }\cdot \xi _{\perp }} \langle P|{\rm Tr}\left[F^{+i}(\xi ^{-},\xi _{\perp })\mathcal{U}%
^{[+]\dagger }F^{+i}(0)\mathcal{U}^{[+]}\right]|P\rangle \ ,\label{g1fund}
\end{eqnarray}%
where the gauge link $\mathcal{U}_\xi^{[+]}=U^n\left[0,+\infty;0\right]U^n%
\left[+\infty, \xi^{-}; \xi_{\perp}\right]$ with $U^n$ being reduced to the light-like Wilson line
in covariant gauge. Within the small-$x$ color glass condensate (CGC) 
framework, this distribution can be written in terms of the
correlator of four Wilson lines as~\cite{Dominguez:2010xd,Dominguez:2011wm},
\begin{equation}
xG^{(1)}(x,k_\perp)=-\frac{2}{\alpha_S}\int\frac{d^2x_\perp}{(2\pi)^2}\frac{d^2y_\perp}{(2\pi)^2}\;
e^{-ik_\perp\cdot(x_\perp-y_\perp)}\left\langle{\rm
Tr}\left[\partial_iU(x_\perp)\right]U^\dagger(y_\perp)
\left[\partial_iU(y_\perp)\right]U^\dagger(x_\perp)\right\rangle_{x},\label{e3}
\end{equation}
where the Wilson line $U(x_{\perp})$ is defined as
$U^n\left[-\infty,+\infty; x_{\perp}\right]$.
The second gluon distribution, the Fourier transform of the dipole
cross section, is defined in the fundamental
representation
\begin{eqnarray}
xG^{(2)}(x,k_{\perp }) &=&2\int \frac{d\xi ^{-}d\xi _{\perp }}{(2\pi
)^{3}P^{+}}e^{ixP^{+}\xi ^{-}-ik_{\perp }\cdot \xi _{\perp }}\langle P|{\rm
Tr}\left[F^{+i}(\xi ^{-},\xi _{\perp })\mathcal{U}^{[-]\dagger }F^{+i}(0)%
\mathcal{U}^{[+]}\right]|P\rangle \ ,  \label{g2}
\end{eqnarray}
where the gauge link $\mathcal{U}_\xi^{[-]}=U^n\left[0,-\infty;0\right]U^n\left[-\infty, \xi^{-};
\xi_{\perp}\right]$ stands for initial state interactions. Due to the gauge link in this gluon
distribution being from $-\infty$ to $+\infty$, naturally this gluon distribution can be related to
the color-dipole cross section evaluated from a dipole of size $r_{\perp}$ scattering on the
nucleus target~\cite{Dominguez:2010xd,Dominguez:2011wm},
\begin{equation}
xG^{(2)}(x,q_\perp)= \frac{q_{\perp }^{2}N_{c}}{2\pi^2 \alpha_s} S_{\perp }\int
\frac{d^2r_\perp}{(2\pi)^2}e^{-iq_\perp\cdot r_\perp} \frac{1}{N_c}\left\langle{\rm
Tr}U(0)U^\dagger(r_\perp)\right\rangle_{x} .\label{e5}
\end{equation}
Similar analysis can also be extended to the polarization dependent
 cases~\cite{Metz:2011wb,Dominguez:2011br,Zhou:2013gsa,Boer:2015pni,Boer:2016xqr}.
Such identifications have laid solid
foundation for the exploration of the nucleon/nuclei tomography in terms of parton distributions, which can be
measured through various high energy hard scattering processes.

 Meanwhile, the QCD evolution
effects also play important roles in describing the scale dependence of these gluon distributions. 
This includes the small-$x$ evolution, i.e., the BFKL/BK
evolution~\cite{Balitsky:1995ub,Kovchegov:1999yj}, and the so-called TMD evolution, i.e., the
Collins-Soper evolution~\cite{Collins:1981uk,Collins:1981uw}. With the small-$x$ approximations
applied in Eq.~(\ref{e3}, \ref{e5}), the small-$x$ evolution effects are taken into account with
the associated evolution equations. However, from those equations, the Collins-Soper evolution
effects are not explicit. A recent quark target model calculation has shown that it is
possible to treat the small $x$ evolution and the TMD evolution in a unified and consistent way by
directly computing the matrix element given in Eq.~(\ref{g1}, \ref{g2}) in the small $x$
limit~\cite{Zhou:2016tfe}. The goal of this paper is to investigate the evolution effects for
these TMDs, by taking into account both small-$x$ and TMD evolution equations
from the perspective of the TMD framework in terms of gauge links. 
Similar studies have been performed by Balitsky and Tarasov in
Refs.~\cite{Balitsky:2015qba} and Marzani in Ref.~\cite{Marzani:2015oyb}. 
Our approach is a bit different from the previous works mentioned above.

We follow closely the derivations in the previous
publications~\cite{Mueller:2012uf,Mueller:2013wwa}, where it has been 
shown the above two resummations (Sudakov and small-$x$) can be
performed consistently at the cross sectoin level. To study the scale dependence of 
TMDs at small $x$, we go back to the full QCD definitions of the TMDs, in which
the scale dependence naturally show up in the associated TMD factorization for 
hard scattering processes. In the gauge
invariant definitions of the gluon distributions as shown in Eqs.~(\ref{g1}, \ref{g2}), there are
un-cancelled light-cone singularities from high order gluon radiations. The regularization introduces
the scheme dependence for the TMDs and the associated factorizations~\footnote{After solving the
evolution equations, the equivalence between different schemes can be proved order by order in
perturbation theory~\cite{Catani:2000vq,Catani:2013tia,Prokudin:2015ysa}.}. 
In our calculation presented in this paper, we follow the Collins 2011
scheme~\cite{collins}, where the soft factor subtraction in the TMDs is applied to regulate the
light-cone singularity. Similar to the case of the hard scattering processes studied in
Refs.~\cite{Mueller:2012uf,Mueller:2013wwa}, the most important high order gluon radiation come
from two regions: (1) soft gluon and (2) collinear gluon. The soft gluon radiation leads to the
Collins-Soper evolution, whereas the collinear gluon contributes to the DGLAP resummation
formulated in terms of the integrated parton distributions in the CSS resummation formalism. In our
case, these collinear gluon radiation contributions actually become the small-$x$ evolution
contributions, which are described by the associated BK/JIMWLK
equations~\cite{Balitsky:1995ub,Kovchegov:1999yj,JalilianMarian:1997jx, Iancu:2003xm}. The above two
contributions are well separated in phase space. That is the reason that we can achieve
resummations of large logarithms from these two sources consistently. The final results for the
TMDs can be written as
\begin{eqnarray}
xG^{(1)}(x,k_\perp,\zeta_c=\mu_F=Q)&=&-\frac{2}{\alpha_S}
\int \frac{d^2x_\perp d^2y_\perp}{(2\pi)^4}e^{ik_\perp\cdot r_\perp}
{\cal H}^{WW}(\alpha_s(Q))e^{-{\cal S}_{sud}(Q^2,r_\perp^2)}\nonumber\\
&&\times {\cal F}^{WW}_{Y=\ln 1/x}(x_\perp,y_\perp)\ ,\label{resum}
\end{eqnarray}
where $r_\perp=x_\perp-y_\perp$, $\zeta_c$ is the regulator for the end-point singularity in the
TMD distributions in the Collins 2011 scheme, $\mu_F$ is the associated factorization scale. In the
final factorization formula, these two scales are usually taken as the same as the hard momentum
scale $Q$ in hard scattering processes.  Meanwhile, ${\cal F}^{WW}_Y$ is the Fourier transform
of the WW gluon distribution as in Eq.~(\ref{e3}),
\begin{equation}
{\cal F}^{WW}_Y(x_\perp,y_\perp)=
\left\langle{\rm Tr}\left[
\partial_\perp^\beta U(x_\perp) U^\dagger(y_\perp)\partial_\perp^\beta
U(y_\perp)U^\dagger(x_\perp)\right]\right\rangle_Y \ .\label{fww}
\end{equation}
and $Y$ represents the rapidity of the gluon from the nucleus $Y\sim \ln (1/x)$. 
The Sudakov form factor contains all order
resummation
\begin{eqnarray}
{\cal S}_{sud}=\int_{c_0^2/r_\perp^2}^{Q^2}\frac{d\mu^2}{\mu^2}\left[A\ln\frac{Q^2}{\mu^2}+B\right] \ ,  \label{sud1}
\end{eqnarray}
where $c_0=2e^{-\gamma_E}$ with $\gamma_E$ the Euler constant.
The hard coefficients $A$ and $B$ can be
calculated perturbatively: $A=\sum\limits_{i=1}^\infty A^{(i)}\left(\frac{\alpha_s}{\pi}\right)^i$ and $B=\sum\limits_{i=1}^\infty B^{(i)}\left(\frac{\alpha_s}{\pi}\right)^i$.
Similarly, we can write dow the result for the dipole-gluon
TMD,
\begin{eqnarray}
xG^{(2)}(x,k_\perp,\zeta_c=\mu_F=Q)&=&-\frac{2}{\alpha_S}
\int \frac{d^2x_\perp d^2y_\perp}{(2\pi)^4}e^{ik_\perp\cdot r_\perp}{\cal H}^{DP}(\alpha_s(Q))e^{-{\cal S}_{sud}(Q^2,r_\perp^2)}\nonumber\\
&&\times {\vec{\bigtriangledown}}^2_{r_\perp}{\cal F}^{DP}_{Y=\ln 1/x}(x_\perp,y_\perp)\
,\label{resum2}
\end{eqnarray}
where  ${\cal F}^{DP}_{Y}(x_\perp,y_\perp)$ with $r_\perp\equiv x_\perp-y_\perp$ is defined as,
\begin{equation}
{\cal F}^{DP}_Y(x_\perp,y_\perp)=\left\langle{\rm Tr}\left[ U(x_\perp)
U^\dagger(y_\perp)\right]\right\rangle_Y \ .\label{fdp}
\end{equation}
In the above equations, both ${\cal F}^{WW}_Y$ and ${\cal F}_Y^{DP}$ are the renormalized quadrupole
and dipole amplitudes, respectively, which obey the associated small-$x$ evolution equations. The TMD evolution
effects are included in the Sudakov factor. The remaining factors, ${\cal H}^{WW}(\alpha_s(Q))$ and ${\cal H}^{DP}(\alpha_s(Q))$, which are of order $1$, are the perturbative calculable finite
hard parts.

The rest of this paper is organized
as follows. In Sec.~II, we present a brief review on the TMD evolution, i.e., the Collins-Soper
evolution, and the CSS resummation\cite{Collins:1984kg} in the collinear framework. Here, the integrated parton
distributions will be important non-perturbative inputs for the TMDs. They obey the DGLAP evolution
equations. In Sec.~III, we compute the TMDs defined in Eqs.(\ref{g1}, \ref{g2}) using the
CGC approach and present our solutions for the TMDs with both Collins-Soper and small-$x$
evolution effects. Finally, we conclude in Sec.~IV.

\section{TMDs in the Collinear Approach and the Collins-Soper evolution }

Before presenting the calculations of TMDs in the CGC approach, it would be instructive to first
briefly review how the Sudakov resummation is formulated in the collinear approach.  We start with
the TMD quark distribution. The un-subtracted TMD quark distribution is defined as
 \begin{eqnarray}
f_q^{unsub.}(x,k_\perp)&=&\frac{1}{2}\int
        \frac{d\xi^-d^2\xi_\perp}{(2\pi)^3}e^{-ix\xi^-P^++i\vec{\xi}_\perp\cdot
        \vec{k}_\perp}  \left\langle
PS\left|\overline\psi(\xi){\cal L}_{n}^\dagger(\xi)\gamma^+{\cal L}_{n}(0)
        \psi(0)\right|PS\right\rangle\ ,\label{tmdun}
\end{eqnarray}
with the gauge link defined as $ {\cal L}_{n}(\xi) \equiv \exp\left(-ig\int^{-\infty}_0 d\lambda
\, v\cdot A(\lambda n +\xi)\right)$. As mentioned in the Introduction,
there exist the light-cone singularities in the un-subtracted TMD
distributions. In the Collins 2011 prescription, these singularities
are cancelled out by the soft factor,
\begin{equation}
\tilde f_{q}^{(sub.)}(x,r_\perp,\mu_F,\zeta_c)=f_q^{unsub.}(x,r_\perp)\sqrt{\frac{S^{\bar
n,v}(r_\perp)}{S^{n,\bar n}(r_\perp)S^{n,v}(r_\perp)}} \ , \label{jcc}
\end{equation}
with $S^{v_1,v_2}$
defined as
\begin{equation}
S^{v_1,v_2}(r_\perp)={\langle 0|{\cal L}_{v_2}^\dagger(r_\perp) {\cal
L}_{v_1}^\dagger(r_\perp){\cal L}_{v_1}(0){\cal
L}_{v_2}(0)  |0\rangle   }\, . \label{softg}
\end{equation}
Here,  $r_\perp$ is the Fourier conjugate variable with respect to the transverse momentum
$k_\perp$, $ \mu_F$ is the factorization scale.  And $\zeta_c^2$ is defined as
$\zeta_c^2=x^2(2v\cdot P)^2/v^2=2(xP^+)^2e^{-2y_n}$ with $y_n$ being the rapidity cut-off in the
Collins-11 scheme. The second factor represents the soft factor subtraction with $n$ and $\bar n$
as the light-front vectors $n=(1^-,0^+,0_\perp)$, $\bar n=(0^-,1^+,0_\perp)$, whereas $v$ is an
off-light-front $v=(v^-,v^+,0_\perp)$ with $v^-\gg v^+$.

\subsection{Soft Gluon Radiation and the Collins-Soper Evolution}

At leading order, the quark TMD in the collinear factorization can be expressed as,
\begin{equation}
f_q(x,k_\perp)|^{(0)}=\delta^{(2)}(k_\perp) q(x) \ ,
\end{equation}
where $q(x)$ represents the integrated quark distribution. In the Fourier
transformation $r_\perp$ space, we have
\begin{equation}
\tilde f_q(x,r_\perp)|^{(0)}= q(x) \ .
\end{equation}
At one-loop order, the most important contribution comes from soft gluon radiation, which leads to
the Collins-Soper evolution equation for the TMDs. To illustrate the energy dependence, it is
convenient to show the one-loop result,
\begin{equation}
f_q(x,k_\perp)|_{soft-real}^{(1)}=\frac{\alpha_s}{2\pi^2}\frac{1}{k_\perp^2}C_F
\int \frac{dx'}{x'}q(x')\left\{\delta(1-\xi)\left(\ln\frac{\zeta_c^2}{k_\perp^2}\right)\right\} \ ,
\end{equation}
where $\xi=x/x'$ and $\zeta_c$ is defined above.
The virtual diagram only contributes to the counter terms,
\begin{eqnarray}
\tilde f_q(x,r_\perp)|_{vir}^{(1)}&=&\int\frac{dx'}{x'}q(x')\frac{\alpha_s}{2\pi}C_F\delta(1-\xi)
\left[-\frac{1}{\epsilon^2}-\frac{3}{2\epsilon}+\frac{1}{\epsilon}\ln\frac{\zeta_c^2}{\mu^2}
\right]  \ ,
\end{eqnarray}
where we have applied the dimensional regulation $d=4-2\epsilon$.
The virtual and soft contributions together give
\begin{eqnarray}
\tilde f_q(x,r_\perp)|_{soft}^{(1)}&=&q(x)\frac{\alpha_s}{2\pi}
C_F\left[\frac{1}{2}\left(\ln\frac{\zeta_c^2}{\mu^2}\right)^2-\frac{1}{2}\left(\ln\frac{\zeta_c^2r_\perp^2}{c_0^2}\right)^2
-\frac{3}{2\epsilon}\right] \, ,
\end{eqnarray}
with $c_0=2e^{-\gamma_E}$. Clearly, the above result demonstrates that we do have energy
dependence.

\subsection{Collinear Gluon Radiation and the DGLAP Evolution}

In addition, there are also collinear gluon radiation contributions, which
can be resummed to all order in the CSS formalism as well. At one-loop order
the collinear gluon takes the following expression,
\begin{equation}
f_q(x,k_\perp)|_{coll-real}^{(1)}=\frac{\alpha_s}{2\pi^2}\frac{1}{k_\perp^2}C_F
\int \frac{dx'}{x'}q(x')\left\{\frac{1+\xi^2}{(1-\xi)_+}+\frac{D-2}{2}(1-\xi)\right\} \ ,
\end{equation}
where $D$ represents the number of transverse dimensions which implies $D=2-2\epsilon$ in the dimensional regulation used in our
calculations. In $r_\perp$ space, it reads,
\begin{eqnarray}
\tilde f_q(x,r_\perp)|_{coll}^{(1)}&=&\frac{\alpha_s}{2\pi}C_F\int
\frac{dx'}{x'}q(x')\left\{\left(-\frac{1}{\epsilon}+\ln\frac{c_0^2}
{r_\perp^2 \mu^2}\right){\cal P}_{qq}(\xi)\right.\nonumber\\
&&\left.+(1-\xi)+\delta(1-\xi)\left[\frac{3}{2}\ln\frac{r_\perp^2\mu^2}{c_0^2}+\frac{3}{2\epsilon}
\right]\right\} \ ,
\end{eqnarray}
where ${\cal P}_{qq}(\xi)=\left(\frac{1+\xi^2}{1-\xi}\right)_+$.
At small-$x$, there is also important contribution from gluon splitting,
which can be written as
\begin{equation}
f_q(x,k_\perp)|_{coll-real}^{(1)}=\frac{\alpha_s}{2\pi^2}\frac{1}{k_\perp^2}T_R
\int \frac{dx'}{x'}g(x')\left\{\left[\xi^2+(1-\xi)^2\right]+(D-2)2\xi(1-\xi)\right\} \ ,
\end{equation}
which contributes to both the integrated quark distribution
and the TMD quark distribution in $r_\perp$-space.

Two evolution equations are derived for the TMDs: one is the energy evolution equation respect to
$\zeta$, the so-called Collins-Soper evolution equation~\cite{Collins:1981uk}; one is the
renormalization group equation associated with the factorization scale $\mu$. After solving the
evolution equations and expressing the TMDs in terms of the integrated parton distributions to have
a complete resummation results, we can write,
\begin{eqnarray}
\tilde f_q^{(sub.)}(x,r_\perp,\zeta_c=\mu_F=Q)&=& e^{-{ {S}_{pert}^q(Q,r_\perp)}}
\widetilde{{\cal F}}_q\left(\alpha_s(Q)\right)\nonumber\\
&&\times \sum_iC_{q/i}(\mu_r/\mu)\otimes f_i(x,\mu)\ , \label{tmdqf}
\end{eqnarray}
where $f_i(x,\mu)$ represent the relevant integrated parton distributions, and
$\mu_r=c_0/r_\perp$. We have chosen the energy parameter $\zeta_c^2=Q^2$ and the factorization scale 
$\mu_F=Q$ to resum large logarithms. The Sudakov factor can be written as
\begin{eqnarray}
S_{pert}^q(Q,r_\perp)=\int_{\mu_r^2}^{Q^2}\frac{d\mu^2}{\mu^2}\left[A_q\ln\frac{Q^2}{\mu^2}+B_q\right] \ ,\label{sudakovq}
\end{eqnarray}
where $A$ and $B$ are perturbative calculable: $A_q=\sum_{i=1}A_q^{(i)}(\alpha_s/\pi)^i$ and
$B_q=\sum_{i=1}B_q^{(i)}(\alpha_s/\pi)^i$, with $A_q^{(1)}=\frac{1}{2}C_F$ and $B_q^{(1)}=-\frac{3}{4}C_F$. 
By choosing the scale $\mu=\mu_r$ for the integrated parton distribution in Eq.~(\ref{tmdqf}), 
the $C$-coefficients are defined as
\begin{equation}
C_{q/q'}(x,\mu=\mu_r)=\delta_{qq'}\left[\delta(1-x)+\frac{\alpha_s}{2\pi}C_F(1-x)\right]\ ,\label{univerc}
\end{equation}
for the quark-quark splitting case. In the Collins-11 (JCC) scheme~\cite{collins}, the scheme
dependent term $\widetilde{\cal F}$ takes a very simple form at one-loop order,
\begin{equation}
\widetilde{\cal F}_q\left(\alpha_s(Q)\right)=1+{\cal O}(\alpha_s^2) \ ,\label{jccf}
\end{equation}
where $\alpha_s$ correction vanishes.

At small-$x$, the dominant contribution comes from the gluon splitting. This
is represented by the collinear quark distribution dependence in Eq.~(\ref{tmdqf}),
where $f_q(x,\mu_r)$ is driven by the gluon distribution at small-$x$ in the
collinear framework. Therefore, at small-$x$, we can find out that the TMD
quark distribution behaves as $\sim g(x,\mu_r)\alpha_s\ln(\mu_r)$.
This is also true in the CGC formalism, which will be discussed in Sec.~III.

\subsection{TMD Gluon Distributions}

Now let us also summarize some known results for the gluon TMD computed in the collinear framework. Again, we follow
the Collins 2011 scheme to define the TMD gluon distributions. The un-subtracted gluon TMD is
defined as
 \begin{eqnarray}
xg^{unsub.}(x,k_\perp,\mu,\zeta,\rho)&=&\int\frac{d\xi^-d^2\xi_\perp}{P^+(2\pi)^3}
    e^{-ixP^+\xi^-+i\vec{k}_\perp\cdot \vec\xi_\perp}\nonumber\\
    &&\times
    \left\langle P|{F_a^+}_\mu(\xi^-,\xi_\perp)
{\cal L}^\dagger_{nab}(\xi^-,\xi_\perp) {\cal L}_{nbc}(0,0_\perp)
F_c^{\mu+}(0)|P \right\rangle\ ,\label{gpdfu}
\end{eqnarray}
where the associated gauge link is in adjoint representation. In the Collins 2011 scheme,
we define
\begin{equation}
xg^{(sub)}(x,b_\perp,\mu,\zeta_c)=xg^{unsub.}(x,b_\perp)\sqrt{\frac{S^{\bar
n,v}(b_\perp)}{S^{n,\bar n}(b_\perp)S^{n,v}(b_\perp)}} \ ,
\end{equation}
where $S^{v_1,v_2}$ are defined similarly as those in the quark
distribution but in the adjoint representations. It is
straightforward to calculate the real correction contribution,
\begin{equation}
f_g^{(sub)}(x,k_\perp)|_{real}^{(1)}=\frac{\alpha_s}{\pi^2}\frac{1}{k_\perp^2}C_A \int
\frac{dx'}{x'}g(x')\left\{ \frac{\xi}{(1-\xi)_+} +
\frac{1-\xi}{\xi}+\xi(1-\xi)+\delta(1-\xi)\left(\ln\frac{\zeta_c^2}{k_\perp^2}\right)\right\} \ .
\end{equation}
In $r_\perp$ space, one has,
\begin{eqnarray}
\tilde f_g^{(sub)}(x,r_\perp)|_{real}^{(1)}&=&\frac{\alpha_s}{2\pi} C_A\int \frac{dx'}{x'}g(x')\left\{
\left( -\frac{1}{\epsilon}+\ln\frac{c_0^2} {r_\perp^2 \mu^2}\right) \left [{\cal
P}_{gg}(\xi)-2\beta_0\delta(1-\xi) \right ] \right \}  \nonumber \\   &+& g(x) \frac{\alpha_s}{2\pi}
C_A \left \{\frac{1}{\epsilon^2}-\frac{1}{\epsilon}\ln\frac{\zeta_c^2}{\mu^2}+
\frac{1}{2}\left(\ln\frac{\zeta_c^2}{\mu^2}\right)^2-
\frac{1}{2}\left(\ln\frac{\zeta_c^2r_\perp^2}{c_0^2}\right)^2 \right \} \ ,
\end{eqnarray}
where ${\cal P}_{gg}(\xi)=\frac{2(1-\xi)}{\xi}+\frac{2\xi}{(1-\xi)_+}+2\xi(1-\xi)-2\beta_0\delta(1-\xi)$
and $\beta_0=\left(11-2N_f/3\right)/12$. After removing  the UV
divergence, the virtual contribution is given by,
\begin{eqnarray}
 \tilde f_g^{(sub)}(x,r_\perp)|_{vir}^{(1)}&=&g(x)\frac{\alpha_s}{2\pi}C_A
\left[-\frac{1}{\epsilon^2}-2\beta_0 \frac{1}{\epsilon}+\frac{1}{\epsilon}\ln\frac{\zeta_c^2}{\mu^2}
\right] \ .
\end{eqnarray}
Combining the real contribution and the virtual contribution, we end up with,
\begin{eqnarray}
\tilde f_g^{(sub)}(x,r_\perp)^{(1)}&=&\frac{\alpha_s}{2\pi}C_A \int \frac{dx'}{x'}g(x')\left\{ \left(
-\frac{1}{\epsilon}+\ln\frac{c_0^2} {r_\perp^2 \mu^2}\right) {\cal P}_{gg}(\xi) \right \} \nonumber
\\   &+& g(x) \frac{\alpha_s}{2\pi} C_A \left
\{\frac{1}{2}\left(\ln\frac{\zeta_c^2}{\mu^2}\right)^2-
\frac{1}{2}\left(\ln\frac{\zeta_c^2r_\perp^2}{c_0^2}\right)^2 +2\beta_0\ln\frac {r_\perp^2
\mu^2}{c_0^2} \right \} \ ,
\end{eqnarray}
where the double and single IR poles are cancelled out. The remaining collinear divergence can be
removed by introducing a renormalized integrated gluon distribution. Following the same procedure
as for the quark TMD case, all large logarithms appear in the above formula can be summed into the
Sudakov factor. All order resummation result can be written, similarly,
\begin{eqnarray}
\tilde f_g^{(sub.)}(x,r_\perp,\zeta_c=\mu_F=Q)&=& e^{-{ {S}_{pert}^g(Q,r_\perp)}}
\widetilde{{\cal F}}_g\left(\alpha_s(Q)\right)\nonumber\\
&&\times \sum_iC_{g/i}(\mu_r/\mu)\otimes f_i(x,\mu)\ , \label{tmdgf}
\end{eqnarray}
where the Sudakov factor takes the same form as in Eq.~(\ref{sudakovq}),
\begin{eqnarray}
S_{pert}^g(Q,r)=\int_{\mu_r^2}^{Q^2}\frac{d\mu^2}{\mu^2}\left[A_g\ln\frac{Q^2}{\mu^2}+B_g\right] \ ,\label{sudakovg}
\end{eqnarray}
 with
$A_g^{(1)}=\frac{1}{2}C_A$ and $B_g^{(1)}=-\beta_0C_A$. Both $C_{g/g}$ and $\widetilde{\cal F}_g$
with $\mu=\mu_r$ in Eq.~(\ref{tmdgf}) vanishes at one-loop order.

At small $x$, we can further simplify  the above expression by approximating ${\cal
P}_{gg}\approx \frac{2}{\xi}$ and treating $x_g' g(x_g')$ as a slowly varying function of $x'$. One
then can trivially carry out $x'$ integration,
\begin{eqnarray}
x\tilde f_g^{(sub)}(x,r_\perp)^{(1)}&=&\frac{\alpha_sC_A}{\pi} xg(x)\ln \frac{1}{x}  \left(
-\frac{1}{\epsilon}+\ln\frac{c_0^2} {r_\perp^2 \mu^2}\right)  \nonumber
\\   &+&x g(x) \frac{\alpha_s}{2\pi} C_A \left
\{\frac{1}{2}\left(\ln\frac{\zeta_c^2}{\mu^2}\right)^2-
\frac{1}{2}\left(\ln\frac{\zeta_c^2r_\perp^2}{c_0^2}\right)^2 +2\beta_0\ln\frac {r_\perp^2
\mu^2}{c_0^2} \right \} \ ,
\end{eqnarray}
where the large logarithm $\ln \frac{1}{x}$ should also be properly resummed  to improve
perturbative calculation. We will address this issue in the next section.

\section{TMDs in the CGC approach and the Collins-Soper evolution}

 It has been argued that at sufficient small $x$ the logarithm $\ln \frac{1}{x}$ is more
important than the collinear logarithm $\ln \frac{Q^2}{\mu^2}$. The leading region, which gives rise to
$\ln \frac{1}{x}$ enhancement, is the so-called strong rapidity ordering region rather than the
strong $k_T$ ordering region, i.e. the leading region in the collinear approach.  As a consequence,
one should keep incoming parton transverse momentum when computing physical observables or parton
TMDs. Note that the Collins-Soper type logarithms can be equally important as the logarithm $\ln
\frac{1}{x}$ depending on kinematics in a physical process. We are aiming to resum these two type
large logarithms simultaneously in a unified and consistent framework. To this end, we formulate
the calculation of parton TMDs in the CGC framework. We  start with discussing  the small $x$ quark
TMDs.

\subsection{TMD  quark at small-$x$}

The quark distribution has been evaluated in the small-$x$ formalism in the
literature~\cite{McLerran:1998nk,Mueller:1999wm,Marquet:2009ca}. It is interesting to note
that the TMD quark distribution for the DIS and Drell-Yan processes are the same, which reflects the
universality of the quark distribution. In this subsection, we review these results, in the context
of the TMD definitions with associated gauge links.

For the Drell-Yan process, the gauge invariant quark distribution contains the gauge link in the
fundamental representation pointing to the direction towards $-\infty$, whereas that for the DIS process the gauge
link goes to $+\infty$. Because of this difference, there are different diagrams that contribute to
the TMD quark distributions for these two processes. We will show that they are the same in the
final result despite the fact that they take the very different forms initially.

\begin{figure}[tbp]
\begin{center}
\includegraphics[height=5cm]{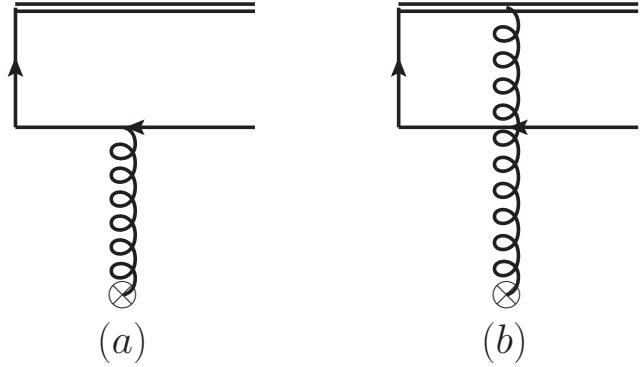}
\end{center}
\caption[*]{The transverse momentum dependent quark distribution calculated in
the small-$x$ formalism for the Drell-Yan process, where the double line represents
the gauge link contributions from the gauge invariant definition of the quark distribution
for this process.}
\label{dyquark}
\end{figure}

The TMD quark in Drell-Yan process can be calculated from the diagrams in
Fig.~\ref{dyquark}, with the following contributions,
\begin{equation}
xq^{(DY)}(x,k_\perp)=\frac{N_c}{8\pi^4}\int d\xi\int
d^2k_{g\perp}F_{x_g}(k_{g\perp})A(k_{g\perp},k_\perp) \ ,
\end{equation}
where $\xi=x/x_g$ with $x_g$ standing for the momentum fraction of the incoming gluon, and $F(q_\perp)$ is
the well-known dipole gluon distribution,
\begin{equation}
F_{x_g}(k_{g\perp})\equiv \int \frac{d^2 x_\perp d^2y_\perp  }{{(2\pi)^2}} e^{-ik_{g\perp} \cdot
(x_\perp-y_\perp)} \frac{1}{N_c}\langle U(x_\perp) U^\dagger(y_\perp)\rangle_{x_g} .
\end{equation}
The coefficient $A$ is defined as
\begin{equation}
A(k_{g\perp},k_\perp) =\left[\frac{\vec{k}_\perp|k_\perp-k_{g\perp}|}{(1-\xi)k_\perp^2+\xi(k_\perp-k_{g\perp})^2}-
\frac{\vec{k}_\perp-\vec{k}_{g\perp}}{|k_\perp-k_{g\perp}|}\right]^{2} \ .
\end{equation}
In the above equation, there is no divergence of the integral over $\xi$, and in the leading
logarithmic small-$x$ approximation, we can integrate out $\xi$ and obtain the following
expression,
\begin{equation}
xq^{(DY)}(x,k_\perp)=\frac{N_c}{4\pi^4}\int d^2k_{g\perp}
F_x(k_{g\perp})\left(1-\frac{k_\perp\cdot(k_\perp-k_{g\perp})}
{k_\perp^2-(k_\perp-k_{g\perp})^2}\ln\frac{k_\perp^2}{(k_\perp-k_{g\perp})^2}\right) \ .
\label{f1q}
\end{equation}
\begin{figure}[tbp]
\begin{center}
\includegraphics[height=5cm]{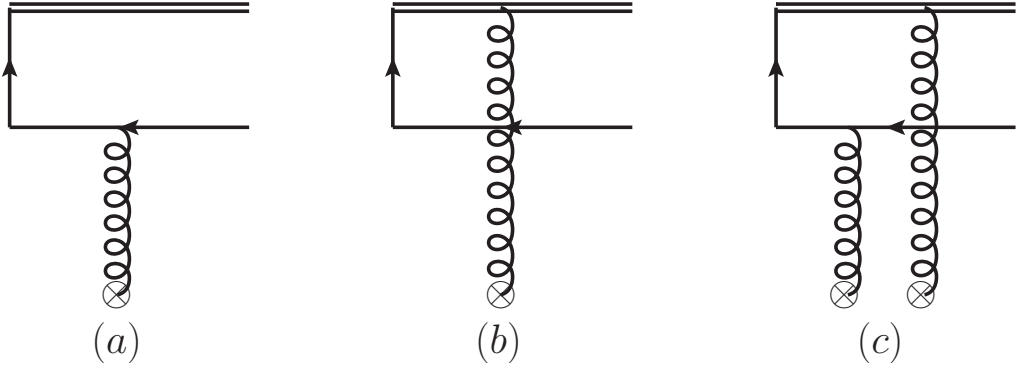}
\end{center}
\caption[*]{Same as Fig.~\ref{dyquark}, but for the DIS process, where
the gauge link goes to $+\infty$.}
\label{disquark}
\end{figure}

However, for the DIS process, because the gauge link goes
to $+\infty$, there is an additional diagram contributing to the
TMD quark distribution, as shown in Fig.~\ref{disquark} (c).
The calculations are straightforward, which gives the following
expression,
\begin{eqnarray}
xq^{(DIS)}(x,k_\perp) &=&  \frac{N_c}{8\pi^4}  \int_0^1  d\xi \int d^2 k_{g\perp}
d^2k_{g1\perp}d^2k_{g1\perp}' \nonumber \\ && \times \int \frac{d^2x_1d^2x_2d^2y_1d^2y_2}{(2\pi)^6}
e^{ik_{g1\perp}\cdot (x_1-x_2)}e^{ik_{g\perp}\cdot (x_2-y_2)}
e^{-ik_{g1\perp}'\cdot (y_1-y_2)}\nonumber\\
&&\times\frac{(k_\perp-k_{g1\perp})\cdot (k_\perp-k_{g1\perp}')(k_\perp-k_{g\perp})^2}
{\left[\xi(k_\perp-k_{g\perp})^2+(1-\xi)(k_\perp-k_{g1\perp})^2\right]
\left[\xi(k_\perp-k_{g\perp})^2+(1-\xi)(k_\perp-k_{g1\perp}')^2\right]}\nonumber\\
&&\times \frac{1}{N_c} \left[\langle U(x_1)U^\dagger(x_2)U(y_2)U^\dagger(y_1)\rangle- \langle
U(x_1)U^\dagger(x_2)\rangle-\langle U(y_2)U^\dagger(y_1)\rangle+1\right].\label{eqm}
\end{eqnarray}
It looks very different as compared to that for the Drell-Yan process. However, one can show that
it is identical to the Drell-Yan quark TMD after a few steps of algebraic manipulations. We start with defining a new variable $\epsilon_f^2=\xi(k_\perp-k_{g\perp})^2/(1-\xi)$ and
 changing integration variable $\xi\rightarrow \epsilon_f^2$.
 The integrand then no longer depends on $k_{g\perp}$. The $k_{g\perp}$ integration thus can be
trivially carried out. As a result, the four point function in the last line collapses into the two
point function after integrating over $y_{2}$. One then arrives at
 \begin{eqnarray}
xq^{(DIS)}(x,k_\perp) &=&  \frac{N_c}{8\pi^4}  \int_0^\infty  d\epsilon_f^2  \int
d^2k_{g1\perp}d^2k_{g1\perp}' \int \frac{d^2x_1d^2x_2d^2y_1}{(2\pi)^4} e^{ik_{g1\perp}\cdot
(x_1-x_2)}
e^{-ik_{g1\perp}'\cdot (y_1-x_2)}\nonumber\\
&&\times\frac{(k_\perp-k_{g1\perp})\cdot (k_\perp-k_{g1\perp}')} {\left[\epsilon_f^2
+(k_\perp-k_{g1\perp})^2\right]
\left[\epsilon_f^2 +(k_\perp-k_{g1\perp}')^2\right]}\nonumber\\
&&\times \frac{1}{N_c} \left\{\langle U(x_1)U^\dagger(y_1)\rangle- \langle
U(x_1)U^\dagger(x_2)\rangle-\langle U(x_2)U^\dagger(y_1)\rangle+1\right\}\ ,
\end{eqnarray}
which can be straightforwardly casted into the following expression
 \begin{eqnarray}
xq^{(DIS)}(x,k_\perp) &=&  \frac{N_c}{8\pi^4} \int_0^\infty  d\epsilon_f^2  \int d^2k_{g\perp} \int
\frac{d^2x_1d^2y_1}{(2\pi)^2} e^{ik_{g\perp}\cdot (x_1-y_1)}
\nonumber\\
&&\times \left [ \frac{\vec k_\perp-\vec k_{g\perp}} {\epsilon_f^2 +(k_\perp-k_{g\perp})^2 }
-\frac{ \vec k_\perp} {\epsilon_f^2 +k_\perp^2}\right ]^2 \frac{1}{N_c} \langle
U(x_1)U^\dagger(y_1)\rangle ,
\end{eqnarray}
with some integration variables renamed. It now becomes evident that the above result for
the DIS quark TMD is the same as that in the Drell-Yan process after changing integration variable
$\epsilon_f^2 \rightarrow \xi$ with $\epsilon_f^2 =\xi (k_\perp-k_{g\perp})^2/(1-\xi)$. As such,
the universality between the DIS quark TMD and the Drell-Yan quark TMD is verified as
expected.

A number of interesting features of this quark distribution have
been discussed in the literature~\cite{{McLerran:1998nk},Mueller:1999wm}.
For example, in the small $k_\perp$ limit, the quark distribution saturates: $
xq(x,k_\perp)|_{k_\perp\to 0}\propto {N_c}/{4\pi^4}$; in the large $k_\perp$ limit,
it has power-law behavior $xq(x,k_\perp)\big |_{k_\perp\gg Q_s}\propto {Q_s^2}/{k_\perp^2}$. Here $Q_s^2$ is the saturation momentum which is proportional to the target gluon density $xG(x)$.
Because of these behaviors, it is legitimate to Fourier transform the above expression into
the $r_\perp$-space,
\begin{equation}
\tilde{f}_{1CGC}^q(x,r_\perp)=\int d^2k_\perp e^{ik_\perp \cdot r_\perp} f_1^q(x,k_\perp) \ ,\label{cgcr}
\end{equation}
where $f_1^q(x,k_\perp)$ follows definition in Eq.~(\ref{f1q}).

Note that both the Collins-Soper type large logarithm and the small $x$ logarithm $\ln \frac{1}{x}$
are absent at leading order.  Beyond the leading order contributions, we anticipate that the higher
order gluon radiation (for instance, a gluon radiated from the gauge link) will generate the Sudakov
logarithms in the soft gluon limit, which can be resummed by solving the associated Collins-Soper
evolution equation. Meanwhile, the next to leading order contribution from the so-called strong rapidity ordering region will give rise to the large logarithm $\ln \frac{1}{x}$ enhancement.  Such contribution should be absorbed into the
renormalized dipole amplitude whose rapidity dependence is governed by the BK equation. After
removing two different type large logarithms by means of the Collins-Soper and the BK equations, respectively, we
are left with finite NLO correction to the hard coefficient.

To implement both TMD and small-$x$ evolution effects, we substitute 
$f_1^q(x,\mu=\mu_r)$ in Eq.~(\ref{tmdqf}) with the
above calculation from the small-$x$ physics,
\begin{equation}
f_1^q(x,\mu_r)\to \widetilde f_{1CGC}^q(x,r_\perp) \ .
\end{equation}
One can justify the above identification (substitution) in the small-$r_\perp$ limit. This is because the integrated
quark distribution at small-$x$ is proportional to the integrated gluon distribution with logarithmic
dependence on the scale $xf_1^q(x,\mu_r)\propto xg(x,\mu_r)\ln (\mu_r)$, which is consistent
with the CGC calculation of $\widetilde f_1^q(x,r_\perp)\propto Q_s^2\ln(1/r_\perp)$.

In the end, we have the following formula for the TMD quark in the CGC formalism,
\begin{equation}
\tilde{f}_q(x,r_\perp,\zeta_c=\mu_F=Q)|_{CGC}= e^{-{S}_{pert}^q(Q,r_\perp)}
\widetilde{f}_{1CGC}^{q}(x,r_\perp)\ ,
\end{equation}
where the Sudakov factor takes the same form as in the last section, Eq.~(\ref{sudakovq}).
In practice, we should implement the non-perturbative part of the Sudakov factor~\cite{collins} as
well.

It is interested to check the small-$r_\perp$, i.e., the $r_\perp\ll 1/Q_s$ behavior, in the above equation. We first notice that
the Fourier transform in Eq.~(\ref{cgcr}) is dominant by the large $k_\perp$ behavior of
$f_1^q(x,k_\perp)$ of Eq.~(\ref{f1q}), and then write
\begin{eqnarray}
\tilde{f}_{1CGC}^q(x,r_\perp)|_{r_\perp\ll 1/Q_s}\propto \int_{Q_s}^{1/r_\perp} d^2k_\perp\frac{Q_s^2}{k_\perp^2}\sim
xG(x,1/r_\perp)\alpha_s\ln(Q_s r_\perp) \ ,
\end{eqnarray}
where we have applied the approximation that the saturation scale $Q_s$ is proportional
to the gluon distribution at small-$x$. We have also 
dropped the exponential factor to simplify the derivation. This is consistent
with the power counting analysis in the last section where $\mu_r=c_0/r_\perp$.

\subsection{TMD gluon at small-$x$}

The TMD gluon distributions can be calculated similarly. The operator definitions given
in the Introductions are the un-subtracted gluon TMDs. In the Collins 2011
scheme, we introduce the same subtraction method as the quark distribution used above
\begin{equation}
f_{g}^{(sub.)}(x,r_\perp,\mu_F,\zeta_c)=f_g^{unsub.}(x,r_\perp)\sqrt{\frac{S^{\bar
n,v}(r_\perp)}{S^{n,\bar n}(r_\perp)S^{n,v}(r_\perp)}} \ , \label{jccg}
\end{equation}
with $S^{v_1,v_2}$ defined in the adjoint representation and $f_g$ representing either of
$G^{(1)}$ and $G^{(2)}$ distributions. The soft gluon contributions are the same for both
distributions. This implies that they obey the same Collins-Soper evolution equation.

We can calculate the above subtracted TMD gluon distributions in the CGC framework. At the leading
order, they are reduced to  the expressions presented in the Introduction, respectively (without
evolution effects). At one-loop order, we again have soft and collinear gluon radiation
contributions. The divergence in the soft gluon radiation is regulated by the soft factor subtraction, which is the same as the collinear approach. The remaining finite part contributes to the Delta function with large
Sudakov logarithms. This part demonstrates that the TMD gluon distributions at small-$x$ obey the
same Collins-Soper evolution equation. To this end, we carry out calculations for the various
TMD gluon distributions at small-$x$ in the CGC framework at one-loop order as follows.

\subsubsection{WW gluon distribution in Higgs boson production process}

\begin{figure}[tbp]
\begin{center}
\includegraphics[height=4cm]{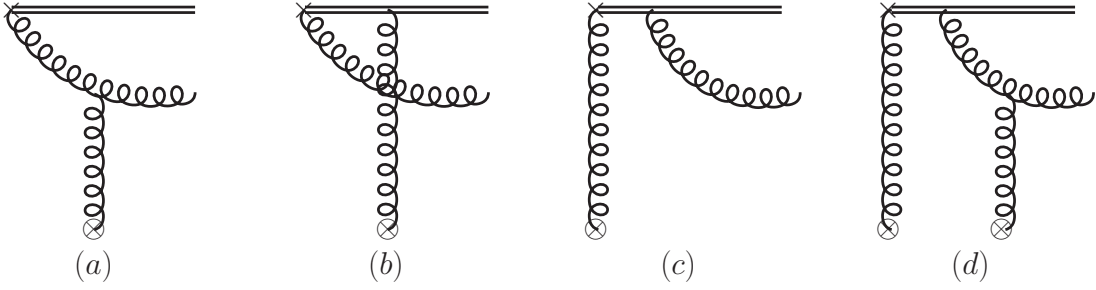}
\end{center}
\caption[*]{The one-loop real gluon radiation contribution
to the TMD gluon distribution with gauge link to $-\infty$ at small-$x$
in the CGC framework. This gluon
distribution can be used to describe the Higgs boson
production in $pA$ collisions. The crosses represent the
probing gluon momentum $(x,k_\perp)$ where $x$
is the longitudinal momentum fraction and $k_\perp$
is the transverse momentum.}
\label{dygluon}
\end{figure}

As a first example, we calculate the WW-gluon distribution in the Higgs boson
production process, where the gauge link in the TMD definition goes to $-\infty$.
At the leading order, the gluon distribution can be written as
\begin{equation}
xG^{(WW)}_{(-\infty)}=\ \frac{-2}{\alpha_s}{\cal F}^{(WW)}(x,k_\perp)=
-\frac{2}{\alpha_s}\int\frac{d^2x_\perp d^2y_\perp}{(2\pi)^{4}}e^{ik_\perp
\cdot(x_\perp-y_\perp)}{\cal F}^{WW}(x_\perp,y_\perp) \ ,
\end{equation}
which is related to the quadruple correlation ${\cal F}^{WW}(x_\perp,y_\perp)$ defined in the
Introduction in the CGC framework. At one-loop order, the Feynman diagrams of real gluon radiation are illustrated in Fig.~\ref{dygluon} with the gauge link going to $-\infty$.

The diagrams in Fig.~\ref{dygluon} (a) and (b) depend on the same Wilson line, which is an adjoint
representation Wilson line from $-\infty$ to $+\infty$. Therefore, we can combine these two
diagrams together,
\begin{eqnarray}
Fig.~\ref{dygluon}(a,b)&\propto& \int
\frac{d^2x_\perp}{(2\pi)^2}e^{ik_{g\perp}\cdot x_\perp}\frac{1}{T_F}{\rm Tr}\left[T^bU(x_\perp)T^a U^\dagger(x_\perp)\right]\nonumber\\
&&\times \frac{1}{2} \left[(1-\xi)\left(\frac{k_\perp^\alpha k_\perp^\nu}{(1-\xi)k_\perp^2+\xi
k_{1\perp}^2}-
\frac{k_{1\perp}^\alpha k_{1\perp}^\nu}{k_{1\perp}^2}\right)\right.\nonumber\\
&&\left.
~~~+\xi k_{1\perp}^2\frac{g_\perp^{\alpha\nu}}{2}\left(\frac{1}{k_{1\perp}^2}-\frac{1}
{(1-\xi)k_\perp^2+\xi k_{1\perp}^2}\right)\right] \ ,
\end{eqnarray}
where $\xi=x/x_g$ is the same as used above, and $k_{g\perp}=k_\perp+k_{1\perp}$ with the radiated gluon transverse
momentum being $k_{1\perp}$. Clearly, in the above result, there is no singularity at the end point
of $\xi\to 1$. That means they do not contribute to the Sudakov logarithms. However, when $\xi\ll
1$, they play important role for the small-$x$ evolution for the WW-gluon distribution.

The end point singularity actually comes from the diagrams in Fig.~\ref{dygluon}(c) and (d). Their
contributions are proportional  to the following expression,
\begin{eqnarray}
Fig.~\ref{dygluon}(c,d)&\propto& \int
\frac{d^2k_{g1\perp}}{(2\pi)^2}\frac{k_\perp^\alpha-k_{g1\perp}^\alpha}{(k_\perp-k_{g1\perp})^2}
\int d^2x_1d^2x_2 e^{ik_{g1\perp}\cdot (x_1-x_2)}e^{ik_{g\perp}\cdot x_2}\nonumber\\
&&\times
{\rm Tr}\left[U^\dagger (x_2)T^bU(x_2)\left[i\partial_\perp^\nu U^\dagger(x_1)U(x_1),T^a\right]\right] \ ,
\end{eqnarray}
which also contributes to the small-$x$ evolution. To see this more
clearly, we can compute the amplitude square of the above term, and
find that it is proportional to
\begin{eqnarray}
\left[Fig.~\ref{dygluon}(c,d)\right]^2&\propto& \frac{1}{(k_\perp-k_{g\perp})^2}
\frac{\alpha_s}{2\pi^2}C_A \int\frac{d\xi}{\xi(1-\xi)}{\cal F}^{(WW)}(x_g,k_{g\perp}) \ .
\end{eqnarray}
The above expression contains two divergent contributions: one with $\xi=1$ and the other with $\xi=0$, which
correspond to the end-point singularity and small-$x$ divergence, respectively. The end-point
singularity of $\xi=1$ will be cancelled by the soft factor subtraction following the Collins 2011
scheme. We would like to emphasize that the soft factor subtraction only applies to the
contribution at the end point of $\xi=1$. Meanwhile, the small-$x$ divergence will be absorbed into
the relevant small-$x$ evolution for the WW-gluon distributions.

We can summarize the total contributions from Fig.~\ref{dygluon} in the following
expression,
\begin{eqnarray}
Fig.~\ref{dygluon}&\propto&\left[\frac{k_{1\perp}^\alpha k_{1\perp}^\nu-\epsilon_f^2\frac{g_\perp^{\alpha\nu}}{2}}{k_{1\perp}^2+\epsilon_f^2}
-\frac{k_{\perp}^\alpha k_{\perp}^\nu-\epsilon_f^2\frac{g_\perp^{\alpha\nu}}{2}}{k_{\perp}^2+\epsilon_f^2}\right]
\Gamma_B(k_{g\perp})+
 \int\frac{d^2k_{g1\perp}}{(2\pi)^2}\frac{k_{1\perp}^{\prime \alpha}}{k_{g1\perp}^{\prime 2}}
\Gamma_A^\nu(k_{g1\perp},k_{g2\perp}) 
\ ,
\end{eqnarray}
where $k_{1\perp}=k_\perp-k_{g\perp}$, $k_{1\perp}'=k_\perp-k_{g1\perp}$, $\epsilon_f^2=\xi k_{1\perp}^2/(1-\xi)$
and $\Gamma_A$ and $\Gamma_B$
are defined as
\begin{eqnarray}
\Gamma_A^\nu&=&\int d^2x_{1\perp}d^2x_{2\perp}e^{ik_{g1\perp}\cdot x_{1\perp}+ik_{g2\perp}\cdot x_{2\perp}}
{\rm Tr}\left[U^\dagger(x_2)T^bU(x_2)[i\partial_\perp^\nu U^\dagger(x_1)U(x_1),T^a]\right] \ ,\nonumber\\
\Gamma_B&=&\int d^2x_\perp e^{ik_{g\perp}\cdot x_\perp} \frac{1}{T_F}{\rm Tr}\left[T^bU(x_\perp)T^aU^\dagger (x_\perp)\right] \ .
\end{eqnarray}
With the above notations, the end point singularity is represented by the limit $\epsilon_f^2\to +\infty$, while the
small-$x$ singularity is obtained by taking $\epsilon_f^2\to 0$. Because the end pint singularity is cancelled out
by the soft factor subtraction, we do not need to consider a particular regulator or cutoff in
the phase space integral.
However, for the small-$x$ part, we have to consider a proper regulator or we have to
impose a kinematic constraint. In this case, because the lower limit for $\xi$ is $x$ of
the probing gluon momentum fraction, the integral limit for $\epsilon_f^2$ will be
around $k_{1\perp}^2/x$. Therefore, the small-$x$ singularity will be represented by
$\ln (1/x)$ in the end. The phase space integral is straightforward, and we obtain the
real gluon radiation contribution at one-loop order,
\begin{eqnarray}
xG^{(WW)}_{(-\infty)}
|_{real}^{(1)}&=&\frac{\alpha_s}{2\pi^2}C_A\left\{\int d^2k_{g\perp}\left(\frac{-2}{\alpha_s}\right){\cal F}^{(WW)}(x,k_{g\perp})
 \frac{1}{(k_\perp-k_{g\perp})^2}\ln\frac{\zeta_c^2}{(k_\perp-k_{g\perp})^2}  \right.\nonumber\\
& +&\left. \ln\left(\frac{1 }{x}\right)\left(\frac{-2}{\alpha_s}\right)\int {\rm\bf K}_{\rm
DMMX(r)}\otimes {\cal F}^{(WW)}(x_g,k_{\perp}) \right\}\ ,
 \end{eqnarray}
where $\zeta_c$ is the regulating parameter in the Collins 2011 scheme and has been defined in
Sec.~II, and  ${\rm\bf K}_{\rm DMMX(r)}$ represents the real part of the BK-type of the small-$x$
evolution kernel\cite{Dominguez:2011gc, Mueller:2013wwa} for the WW-gluon distribution. We would like to emphasize that the final results
with the Sudakov resummation will not depend on how we regulate the end-point singularity as shown
in Refs.~\cite{Catani:2000vq,Catani:2013tia,Prokudin:2015ysa}.

\begin{figure}[tbp]
\begin{center}
\includegraphics[height=4cm]{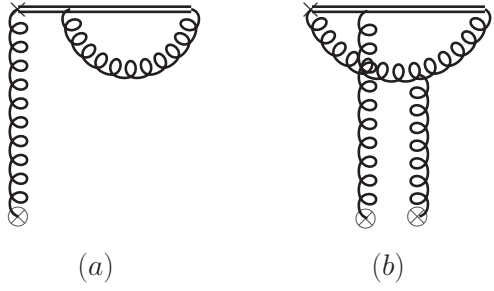}
\end{center}
\caption[*]{The one-loop virtual gluon radiation contribution
to the WW-type of TMD gluon distribution. The virtual diagrams
are the same for the Drell-Yan type and DIS type of gluon distributions.
Again, the crosses represent the
probing gluon momentum $(x,k_\perp)$ where $x$
is the longitudinal momentum fraction and $k_\perp$
is the transverse momentum.}
\label{wwvirtual}
\end{figure}

We can also calculate the virtual graphs, which have been shown in Fig.~\ref{wwvirtual},
where Fig.~\ref{wwvirtual}(a) contributes to both end-point singularity and
small-$x$ divergence and Fig.~\ref{wwvirtual}(b) only contributes
to the small-$x$ divergence.
In particular, the contribution from Fig.~\ref{wwvirtual}(a) can be
written as
\begin{eqnarray}
\frac{\alpha_s}{2\pi}C_A{\cal F}^{(WW)}(x,k_\perp)\int_0^1
\frac{d\xi}{\xi(1-\xi)}\frac{d^2q_\perp}{(2\pi)^2}\frac{1}{q_\perp^2} \ ,
\end{eqnarray}
where we can clearly separate out the above two contributions.
For the end-point singularity term, we subtract it and put it in the soft factor
from the Collins 2011 scheme, and obtain the same result
as that in the collinear framework. Combining with the real
gluon radiation contribution, we find that the finite end-point
contribution leads to the Sudakov double logarithms.

It is a little bit more involved in the calculation of Fig.~\ref{wwvirtual}(b),
for which we obtain,
\begin{eqnarray}
\frac{\alpha_s}{2\pi}C_A\int_0^\infty\frac{d\xi}{\xi}\frac{d^2q_\perp}{(2\pi)^2}\frac{q_\perp\cdot (q_\perp-k_{g2\perp})(q_\perp^\nu-k_{g2\perp}^\nu)-q_\perp^\nu \epsilon_f^{\prime 2}/2}{q_\perp^2((q_\perp-k_{g2\perp})^2+\epsilon_f^{\prime 2})}\Gamma_{AB}^{\prime \nu}(k_\perp,k_{g2\perp}) \ ,
\end{eqnarray}
where $\epsilon_f^{\prime 2}=q_\perp^2/\xi$ and $\Gamma_{AB}'$ is defined as
\begin{eqnarray}
\Gamma_{AB}'&=&\int d^2x_1d^2x_2d^2x_\perp'
e^{ik_{g\perp}\cdot (x_{2\perp}-x_\perp')+ik_{g1\perp}\cdot (x_{1\perp}-x_{2\perp})}\nonumber\\
&&\times \frac{1}{2}\left({\rm Tr}[U^\dagger(x_2)U(x_{1\perp})]{\rm Tr}[U^\dagger(x_{1\perp})U(x_2)i\partial_\perp^\nu U^\dagger(x_\perp')U(x_\perp')]
+h.c.\right)  \ .\label{gammaABP}
\end{eqnarray}
It is important to notice that there is only small-$x$ divergence when $\epsilon_f'=0$. At this
point, it is worthwhile to mention that the above results have been obtained by computing
 the virtual diagrams in the light
cone gauge($A^-=0$) with the principal value prescription. One can also redo the calculation with
other prescription and reproduce the same Collins-Soper and the BK evolution kernels.

Combining the real and virtual contributions, we can write down in the
Fourier transformation conjugate variable $b_\perp$ as
in Sec.II,
\begin{eqnarray}
xG^{(WW)}_{(-\infty)}(x,r_\perp)
|^{(1)}&=&\frac{\alpha_s}{2\pi}C_A\left\{\left(\frac{-2}{\alpha_s}\right)
{\cal F}^{(WW)}(r_\perp)\left[\frac{1}{2}\left(\ln\frac{\zeta_c^2}{\mu^2}\right)^2-\frac{1}{2}\left(\ln\frac{\zeta_c^2r_\perp^2}{c_0^2}\right)^2\right]\right.\nonumber\\
&&\left. +\ln\left(\frac{1}{x}\right)\left(\frac{-2}{\alpha_s}\right)\int {\rm\bf K}_{\rm
DMMX}\otimes {\cal F}^{(WW)}(x_g,r_{\perp}) \right\}\ ,
 \end{eqnarray}
where the finite terms combine the real gluon radiation and 
virtual contributions.
With the above one-loop result, we can derive the Collins-Soper evolution
equation for the TMD gluon. By solving the associated equations, we will be
able to resum the Sudakov logarithms. Similarly, we can derive the small-$x$
evolution, which resums the small-$x$ logarithms.

With all order resummation we obtain the final results for the TMD
gluon distribution as mentioned in the Introduction with Eq.(6).
\begin{eqnarray}
xG^{(WW)}_{(-\infty)}(x,r_\perp,\zeta_c^2=Q^2)&=&-\frac{2}{\alpha_S}
\int \frac{d^2x_\perp d^2y_\perp}{(2\pi)^{4}}e^{ik_\perp\cdot r_\perp}{\cal H}^{WW}(\alpha_s(Q))e^{-{\cal S}_{pert}^g(Q^2,r_\perp^2)}\nonumber\\
&&\times {\cal F}^{WW}_{Y=\ln 1/x}(x_\perp,y_\perp)\ ,\label{resumww}
\end{eqnarray}
where  ${\cal F}^{WW}_{Y=\ln(1/x)}$ is renormalized quadrupole gluon distributions defined in
Eq.~(\ref{fww}), and the Sudakov factor takes the same form as Eq.~(\ref{sudakovg}) in the last
section, except now $B_g^{(1)}=0$. The hard coefficient vanishes at one-loop order too: ${\cal
H}^{WW}=1+{\cal O}(\alpha_s)$.

The result of $B_g^{(1)}=0$ reflects the fact that the running of $\alpha_s$
is treated differently in the small-$x$ CGC formalism than that in
the collinear factorization formalism, where $B_g^{(1)}$ is proportional
to $\beta_0$. In particular, $\alpha_s$ running is part of the
next-to-leading order BK evolution. However, in the collinear
factorization, $\alpha_s$ running effects enters at the one-loop
order (leading order in the DGLAP evolution). It is  also related
to the anomalous dimension for the integrated gluon distribution.
Since we factorize the gluon TMD in terms of the integrated gluon
distribution, we obtain $B_g^{(1)}$ as correspondent term from
the anomalous dimension of the integrated gluon distribution.
In the small-$x$ calculations, however, the TMD gluon is calculated
directly in the CGC formalism. There is no corresponding term
related to the anomalous dimension with the integrated gluon
distribution.

\subsubsection{WW gluon distribution in DIS process}

\begin{figure}[tbp]
\begin{center}
\includegraphics[height=4cm]{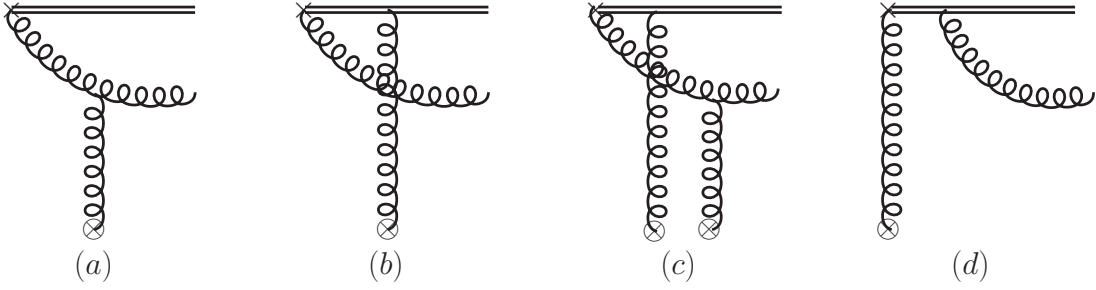}
\end{center}
\caption[*]{The same as Fig.~\ref{dygluon} but with the
gauge link to $+\infty$. This gluon
distribution can be used to describe the semi-inclusive
process in $eA$ collisions, such as the di-jet correlation
in DIS process. Again, the crosses represent the
probing gluon momentum $(x,k_\perp)$ where $x$
is the longitudinal momentum fraction and $k_\perp$
is the transverse momentum.}
\label{disgluon}
\end{figure}

Now let us consider the WW gluon distribution in DIS type of processes\cite{Kovchegov:1998bi, Xie:2013cba},
which only have final state interactions at LO.  
Because of the difference in the gauge link direction, there will be different
diagrams contributing to the WW gluon distribution in the DIS process.
We show the Feynman diagrams for this process in Fig.~\ref{disgluon}.

For this case, it is the diagrams in Fig.~\ref{disgluon} (a), (b) and (c)
contribute to the terms depending on the adjoint representation
Wilson line from $-\infty$ to $+\infty$, while the digram
Fig.~\ref{disgluon} (d) contributes to the end-point
singularity. The calculations of these diagrams follow
those for Fig.~\ref{dygluon}. The total
contributions can be written as
\begin{eqnarray}
Fig.~\ref{disgluon}&\propto&\frac{k_{1\perp}^{\alpha}}{k_{1\perp}^{2}}
\Gamma_A^{\prime \nu}(k_{g\perp})+ \int\frac{d^2k_{g1\perp}}{(2\pi)^2}
\left[\frac{k_{1\perp}^{\prime \alpha} k_{1\perp}^{\prime \nu}-\epsilon_f^2\frac{g_\perp^{\alpha\nu}}{2}}{k_{1\perp}^{\prime 2}+\epsilon_f^2}
-\frac{k_{\perp}^\alpha k_{\perp}^\nu-\epsilon_f^2\frac{g_\perp^{\alpha\nu}}{2}}{k_{\perp}^2+\epsilon_f^2}\right]
\Gamma_B'(k_{g1\perp},k_{g2\perp})  
\ ,
\end{eqnarray}
where again $k_{1\perp}=k_\perp-k_{g\perp}$, $k_{1\perp}'=k_\perp-k_{g1\perp}$, $\epsilon_f^2=\xi k_{1\perp}^2/(1-\xi)$
and $\Gamma_A'$ and $\Gamma_B'$
are defined as
\begin{eqnarray}
\Gamma_A^{\prime \nu}&=&\int d^2x_\perp e^{ik_{g\perp}\cdot x_\perp}
{\rm Tr}\left[T^b[i\partial_\perp^\nu U^\dagger(x_\perp)U(x_\perp),T^a]\right] \ ,\nonumber\\
\Gamma_B'&=&\int d^2x_{1\perp}d^2x_{2\perp}e^{ik_{g1\perp}\cdot x_{1\perp}+ik_{g2\perp}\cdot x_{2\perp}} \frac{1}{T_F}{\rm Tr}\left[U^\dagger(x_2)T^bU(x_2)U^\dagger(x_{1\perp})T^aU (x_{1\perp})\right] \ .
\end{eqnarray}
The phase space integral for the real gluon radiations can be performed following the
previous case, and we obtain the following result,
\begin{eqnarray}
xG^{(WW)}_{(+\infty)}
|_{real}^{(1)}&=&\frac{\alpha_s}{2\pi^2}C_A\left\{\int d^2k_{g\perp}\left(\frac{-2}{\alpha_s}\right){\cal F}^{(WW)}(x,k_{g\perp})
 \frac{1}{(k_\perp-k_{g\perp})^2}\ln\frac{\zeta_c^2}{(k_\perp-k_{g\perp})^2}  \right.\nonumber\\
& +&\left.\ln\left(\frac{1}{x}\right)\left(\frac{-2}{\alpha_s}\right)\int {\rm\bf K}_{\rm
DMMX(r)}\otimes {\cal F}^{(WW)}(x_g,k_{\perp})  \right\}\ .
 \end{eqnarray}
Clearly, the Sudakov term and the small-$x$ divergent term are the same
as that in previous case.

As mentioned in the previous subsection, we will have the same virtual contribution to the DIS-type
gluon TMD. By adding the real and virtual contributions together, we can obtain the similar
expression for the total result at one-loop order with the same Collins-Soper evolution and
small-$x$ evolution terms. Therefore, the resummation for the DIS-type gluon distribution is the
same as that for the Drell-Yan type of gluon TMDs. Moreover, if one does not impose any lower cut
off for $\xi$ integration, one can show that the Drell-Yan like gluon TMD and the DIS gluon TMD are
identical up to finite terms by playing the same mathematica trick applied to the quark TMD case.
This demonstrates that our small $x$ calculation is consistent with the universality argument.
Therefore, after resummation, we have the same WW gluon distribution for the DIS-type and DY-type of processes.

\subsubsection{Dipole gluon distribution}

\begin{figure}[tbp]
\begin{center}
\includegraphics[height=4cm]{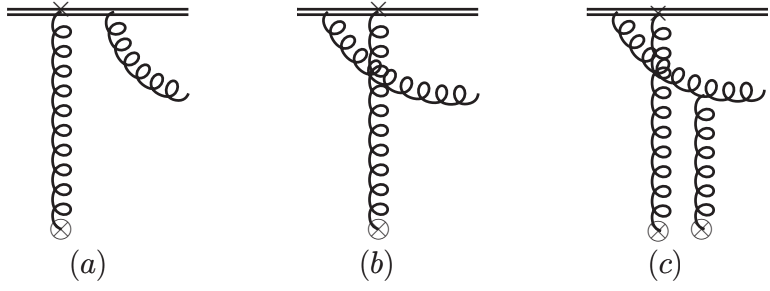}
\end{center}
\caption[*]{Part of one-loop real gluon radiation
diagrams for the dipole gluon distributions contributing
to the Sudakov logarithms.  This gluon
distribution can be used to describe the photon-jet
correlation in $pA$ collisions. Again, the crosses represent the
probing gluon momentum $(x,k_\perp)$ where $x$
is the longitudinal momentum fraction and $k_\perp$
is the transverse momentum.}
\label{dpgluon}
\end{figure}

\begin{figure}[tbp]
\begin{center}
\includegraphics[height=4cm]{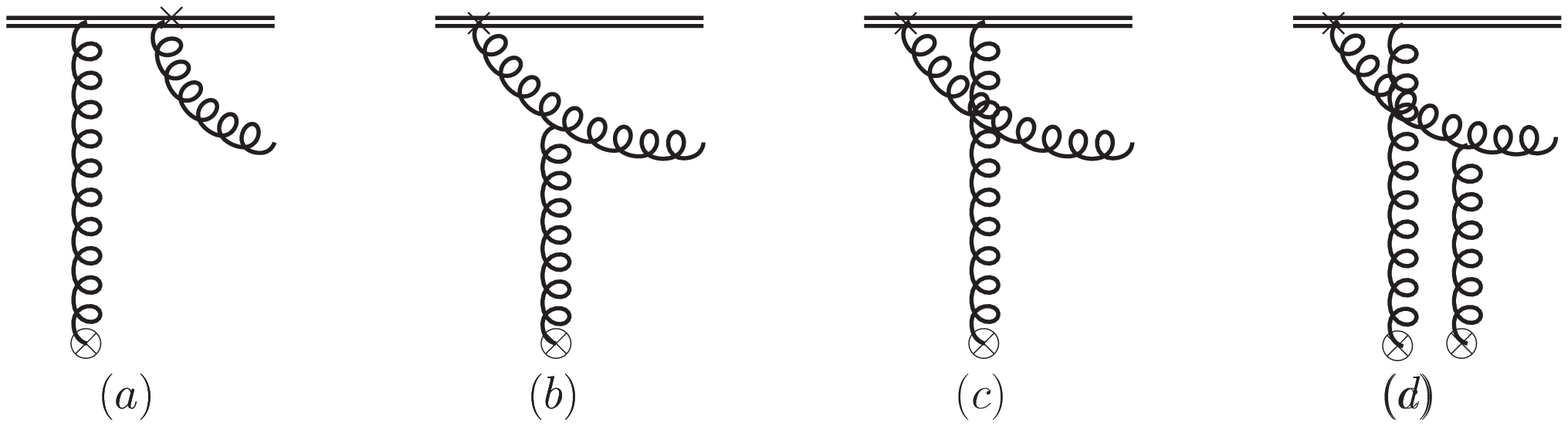}
\end{center}
\caption[*]{Same as Fig.~\ref{dpgluon} but for part of
BK-evolution contributions. Again, the crosses represent the
probing gluon momentum $(x,k_\perp)$ where $x$
is the longitudinal momentum fraction and $k_\perp$
is the transverse momentum.}
\label{dpgluonBK}
\end{figure}

Now, we turn to the dipole gluon distribution. The gauge link structure
is very different from the previous cases. The leading order
result can be expressed as
\begin{equation}
xG^{(dp)}(x,k_\perp)=\frac{-2}{\alpha_s}{\cal F}^{(DP)}(x,k_\perp)=
-\frac{2}{\alpha_s}\int\frac{d^2x_\perp d^2y_\perp}{(2\pi)^{4} } e^{ik_\perp \cdot
(x_\perp-y_\perp)} {\vec{\bigtriangledown}}^2_{r_\perp} {\cal F}^{(DP)}(x_\perp,y_\perp) \ ,
\end{equation}
where the dipole amplitude ${\cal F}^{(DP)}$ has been defined
in the Introduction.
In Fig.~\ref{dpgluon} and \ref{dpgluonBK},
we show the real gluon radiation at one-loop order.

The diagrams in Fig.~\ref{dpgluon} contribute to the Sudakov
logarithms, and we get the following expression,
\begin{eqnarray}
Fig.~\ref{dpgluon}&\propto&
 \int\frac{d^2x_\perp}{(2\pi)^2}e^{ik_{g\perp}\cdot x_\perp}
\frac{k_{1\perp}^\alpha}{k_{1\perp}^2} \left[T^a\partial_\perp^\nu U(x_\perp)\right] \nonumber\\
&&-\int\frac{{d^2k_{g1\perp}}  d^2x_1d^2x_2}{(2\pi)^4}e^{ik_{g1\perp}\cdot (x_1-x_2)}e^{ik_{g\perp}\cdot x_2}
\frac{k_\perp^\alpha-k_{g1\perp}^\alpha}{(k_\perp-k_{g1\perp})^2}
\left[\partial_\perp^\nu U(x_1)U^\dagger(x_2)T^aU(x_2)\right] \ .\label{e40}
\end{eqnarray}
The amplitude squared of the above contribution can be calculated accordingly,
\begin{eqnarray}
[Fig.6]^2 &\propto & 2 \int \frac{d\xi}{\xi (1-\xi)} \left \{ \frac{d^2 x_\perp}{(2\pi)^2}
e^{ik_{g\perp} \cdot (x_\perp-y_\perp)} \frac{1}{k_{1\perp}^2}C_F [\partial_\perp^\mu
U(x_\perp)\partial_\perp^\mu U^\dag(y_\perp)] \right .\ \nonumber
\\  &&- \left .\ \int \frac{d^2k_{g1\perp} d^2 x_{\perp}d^2 y_{1\perp}d^2 y_{2\perp} } {(2\pi)^4} e^{ik_{g\perp} \cdot
(x_{\perp}-y_{2\perp}) }e^{-ik_{g1\perp} \cdot ( y_{1\perp}- y_{2\perp}) } \right .\ \nonumber \\
&&\left .\ \times \frac{k_{1\perp} \cdot k_{1\perp}'}{k_{1\perp}^2 k_{1\perp}'^2} [T^a
\partial_\perp^\mu U(x_\perp) U^\dag(y_{2\perp}) T^a U(y_{2\perp})
\partial_\perp^\mu U^\dag(y_{1\perp})] \right \} \ .
\end{eqnarray}
with $k_{1\perp}'=k_\perp-k_{g1\perp}$. By employing the identity,
\begin{eqnarray}
\textrm{Tr}[T^a \partial_\perp^\mu U(x_\perp) U^\dag(y_{2\perp}) T^a U(y_{2\perp}) \partial_\perp^\mu
U^\dag(y_{1\perp})]&=&\frac{1}{2}\textrm{Tr}[\partial_\perp^\mu U(x_\perp)U^\dag(y_{2\perp})] \textrm{Tr}[U(y_{2\perp})
\partial_\perp^\mu U^\dag(y_{1\perp})]
\nonumber \\ &&-\frac{1}{2N_c}\textrm{Tr}[\partial_\perp^\mu U(x_\perp) \partial_\perp^\mu U^\dag(y_{1\perp})] \ ,
\end{eqnarray}
one  obtains,
\begin{eqnarray}
[Fig.6]^2 &\propto &  \int\frac{d\xi}{\xi (1-\xi)}  \left \{ \frac{d^2 x_\perp}{(2\pi)^2}
e^{ik_{g\perp} \cdot (x_\perp-y_\perp)} \frac{1}{k_{1\perp}^2}2C_A\textrm{Tr}[\partial_\perp^\mu
U(x_\perp)\partial_\perp^\mu U^\dag(y_\perp)] \right .\ \nonumber
\\ &-& \left .\   \int\frac{d\xi}{\xi (1-\xi)} \int \frac{  d^2k_{g1\perp} d^2 x_{\perp}d^2 y_{1\perp}d^2 y_{2\perp} } {(2\pi)^4} e^{ik_{g\perp} \cdot
(x_{\perp}-y_{2\perp}) }e^{-ik_{g1\perp} \cdot ( y_{1\perp}- y_{2\perp}) } \right .\ \nonumber \\
&&\left .\ \times \frac{k_{1\perp} \cdot k_{1\perp}'}{k_{1\perp}^2 k_{1\perp}'^2}
\textrm{Tr}[\partial_\perp^\mu U(x_\perp)U^\dag(y_{2\perp})] \textrm{Tr} [U(y_{2\perp})
\partial_\perp^\mu U^\dag(y_{1\perp})]\right \} \ .
\end{eqnarray}
After the end point singularity in the first term is regularized by subtracting the soft factor in
the Collins-11 scheme, the desired Sudakov logarithm can be recovered.  The second term is a bit
troublesome as the end point singularity is associated with three point function. To get rid of
this contribution, our argument goes as follows. First of all an additional hard scale is always
required for justifying the use of TMD factorization. In the current case, the radiated gluon
transverse momentum plays the role of the hard scale, which implies $k_\perp \gg k_{g\perp} \sim
k_{g1\perp}$.  When we ignore $ k_{g\perp}$ and $ k_{g1\perp}$ in the second term, we can carry out
some integrations trivially and reduce the three point function to the two point function
$\textrm{Tr}[\partial_\perp^\mu U(x_\perp)U^\dag(y_{2\perp})]\textrm{Tr} [U(y_{2\perp})
\partial_\perp^\mu U^\dag(y_{2\perp})]$ which apparently vanishes.

Besides, the above term also contributes to the small-$x$ divergence.
Additional diagrams are shown in Fig.~\ref{dpgluonBK}, which can be
written as
\begin{eqnarray}
Fig.~\ref{dpgluonBK}&\propto&
\int
\frac{d^2x_{1\perp}d^2x_{2\perp}d^2k_{g1\perp}}{(2\pi)^4}e^{ik_{g1\perp}\cdot (x_{1\perp}-x_{2\perp})}e^{ik_{g\perp}\cdot x_{2\perp}} \left[ U(x_1)U^\dagger(x_2)T^aU(x_2)\right]
\nonumber\\
&&\times \frac{1}{2}\left[\frac{k_{1\perp}^{\prime \alpha} k_{1\perp}^{\prime \nu}-\epsilon_f^2\frac{g_\perp^{\alpha\nu}}{2}}{k_{1\perp}^{\prime 2}+\epsilon_f^2}
-\frac{k_{1\perp}^\alpha k_{1\perp}^\nu-\epsilon_f^2\frac{g_\perp^{\alpha\nu}}{2}}{k_{1\perp}^2+\epsilon_f^2}\right] \ ,\label{e41}
\end{eqnarray}
where again $k_{1\perp}=k_\perp-k_{g\perp}$, $k_{1\perp}'=k_\perp-k_{g1\perp}$, $\epsilon_f^2=\xi
k_{1\perp}^2/(1-\xi)$. To check the BK-evolution, we perform the partial integral of
Eq.~(\ref{e40}), and take $\xi=0$ of Eq.~(\ref{e41}), and add them together,
\begin{eqnarray}
Fig.~(\ref{dpgluon},\ref{dpgluonBK})|_{BK}&\propto& 
k_\perp^\nu  \int\frac{d^2x_\perp}{(2\pi)^2}e^{ik_{g\perp}\cdot x_\perp}
\frac{k_{1\perp}^\alpha}{k_{1\perp}^2} \left[T^aU(x_\perp)\right] \nonumber \\
&&- k_\perp^\nu \int\frac{d^2 k_{g1\perp}d^2x_1d^2x_2}{(2\pi)^4}e^{ik_{g1\perp}\cdot (x_1-x_2)}e^{ik_{g\perp}\cdot x_2}
\frac{ (k_\perp-k_{g1\perp})^\alpha}{(k_\perp-k_{g1\perp})^2}   \nonumber \\
&&   \times \left[ U(x_1)U^\dagger(x_2)T^aU(x_2)\right] \ ,
\end{eqnarray}
where the bracket in the above equation is exactly the real gluon contribution to the BK-evolution
for the dipole scattering amplitude.

We can also carry out the phase space integral, and arrive at the following result
for the real gluon radiation contribution to the dipole-gluon distribution at one-loop order,
\begin{eqnarray}
xG^{(DP)}
|_{real}^{(1)}&=&\frac{\alpha_s}{2\pi^2}C_A\left\{\int d^2k_{g\perp}\left(-\frac{2}{\alpha_s}\right){\cal F}^{(DP)}(x,k_{g\perp})
 \frac{k_{g\perp}^2}{(k_\perp-k_{g\perp})^2}\ln\frac{\zeta_c^2}{(k_\perp-k_{g\perp})^2}  \right.\nonumber\\
& +&\left. \ln\left(\frac{1}{x}\right)\left(-\frac{2}{\alpha_s}\right)k_\perp^2\int {\rm\bf
BK}_{\rm (r)}\otimes {\cal F}^{(DP)}(x,k_{g\perp}) \right\}\ .
 \end{eqnarray}

\begin{figure}[tbp]
\begin{center}
\includegraphics[height=4cm]{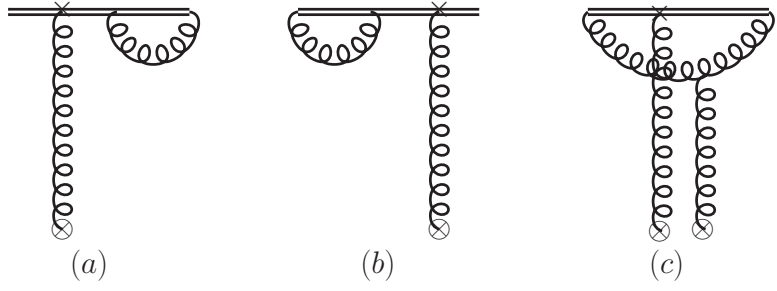}
\end{center}
\caption[*]{Virtual gluon radiation diagrams for
the dipole gluon distribution at one-loop order.
These diagrams contribute to the Sudakov
and small-$x$ logarithms.
Again, the crosses represent the
probing gluon momentum $(x,k_\perp)$ where $x$
is the longitudinal momentum fraction and $k_\perp$
is the transverse momentum.}
\label{dpgluonvir}
\end{figure}

\begin{figure}[tbp]
\begin{center}
\includegraphics[height=4cm]{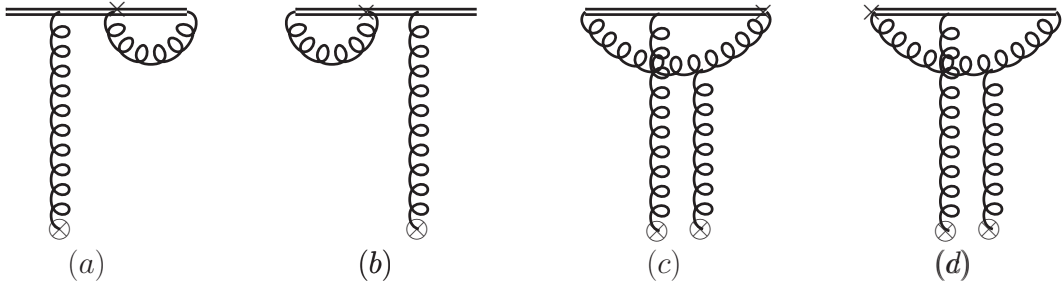}
\end{center}
\caption[*]{Same as Fig.~\ref{dpgluonvir} but for part of
BK-evolution contributions. Again, the crosses represent the
probing gluon momentum $(x,k_\perp)$ where $x$
is the longitudinal momentum fraction and $k_\perp$
is the transverse momentum.}
\label{dpgluonBKvir}
\end{figure}

Similarly, we can calculate the virtual diagrams, which are shown in Figs.~\ref{dpgluonvir},\ref{dpgluonBKvir}.
The calculation of diagrams of Fig.~\ref{dpgluonvir}(a,b)  is the same as that in Fig.~\ref{wwvirtual}(a).
Their contributions, together with the sub-leading $N_c$ contribution from Fig.~\ref{dpgluonvir}(c),
lead to the following expression,
\begin{eqnarray}
\frac{\alpha_s}{2\pi}C_A\int_0^1\frac{dz}{z(1-z)}\frac{d^2q_\perp}{(2\pi)^2}\frac{1}{q_\perp^2} \ .
\end{eqnarray}
The leading $N_c$ contribution from Fig.~\ref{dpgluonvir}(c) leads to
\begin{eqnarray}
\frac{\alpha_s}{2\pi}C_A\int\frac{dz}{z(1-z)}\frac{q_\perp\cdot (q_\perp-k_{g2\perp})}{q_\perp^2(q_\perp-k_{g2\perp})^2}k_{g1\perp}^\nu k_\perp^\nu \Gamma_{AB}^{(dp)} \ ,
\end{eqnarray}
where $\Gamma_{AB}^{(dp)}$ is defined as
\begin{eqnarray}
\Gamma_{AB}^{(dp)}&=& \int_0^1 d^2x_1d^2x_2d^2x_{\perp}'
e^{ik_{\perp}\cdot (x_{2\perp}-x_{\perp}')+ik_{g1\perp}\cdot (x_{1\perp}-x_{2\perp})}\nonumber\\
&&\times {\rm Tr}[U(x_2)U^\dagger(x_\perp')]{\rm Tr}[U(x_1)U^\dagger(x_2)]   \ . \label{gammaABdp}
\end{eqnarray}
It is interesting to find out that the diagrams of Fig.~\ref{dpgluonBKvir}(a,b) vanish,
and the calculations of Fig.~\ref{dpgluonBKvir}(c,d) are similar to those in Fig.~\ref{wwvirtual}(c).
We will have the following expression,
\begin{eqnarray}
\frac{\alpha_s}{2\pi}C_A\int_0^\infty\frac{d\xi}{\xi}\frac{d^2q_\perp}{(2\pi)^2}\frac{q_\perp\cdot (q_\perp-k_{g2\perp})(q_\perp^\nu-k_{g2\perp}^\nu)-q_\perp^\nu \epsilon_f^{\prime 2}/2}{q_\perp^2((q_\perp-k_{g2\perp})^2+\epsilon_f^{\prime 2})} k_\perp^\nu\Gamma_{AB}^{(dp)}(k_\perp,k_{g2\perp}) \ ,
\end{eqnarray}
where $\epsilon_f^{\prime 2}=q_\perp^2/\xi$ and $\Gamma_{AB}^{(dp)}$ is defined as Eq.~(\ref{gammaABdp}).
The above equation only contributes to the small-$x$ divergence. The total
contributions for the small-$x$ divergence from the above equations can be summarized as,
\begin{eqnarray}
k_\perp^\nu k_\perp^\nu\frac{\alpha_s}{2\pi}\int\frac{dz}{1-z}\frac{d^2q_\perp}{(2\pi)^2}\left[\frac{1}{q_\perp^2}-\frac{q_\perp\cdot (q_\perp-k_{g2\perp})}{q_\perp^2(q_\perp-k_{g2\perp})^2} \right]\Gamma_{AB}^{(dp)}(k_\perp,k_{g2\perp}) \ ,
\end{eqnarray}
which is the virtual contribution to the BK evolution of the dipole amplitude.

Adding the real and virtual contributions together, we have the total contribution
at one-loop order,
\begin{eqnarray}
xG^{(DP)}(x,r_\perp) |^{(1)}&=&\frac{\alpha_s}{2\pi}C_A\left\{\left(-\frac{2}{\alpha_s}\right)
\vec{\bigtriangledown}^2_{r_\perp} {\cal F}^{(DP)}(x,r_\perp)\left[
\frac{1}{2}\left(\ln\frac{\zeta_c^2}{\mu^2}\right)^2-\frac{1}{2}\left(\ln\frac{\zeta_c^2r_\perp^2}{c_0^2}\right)^2\right]  \right.\nonumber\\
&&\left.+\ln\left(\frac{1}{x}
\right)\left(-\frac{2}{\alpha_s}\right)\vec{\bigtriangledown}^2_{r_\perp}\int
{\rm\bf K}_{\rm BK}\otimes {\cal F}^{(DP)}(x,r_{\perp})  \right\}\ .
 \end{eqnarray}
The above result is almost the same as that in previous subsection for the $WW$-gluon
distribution. Again, we have double logarithms associated with Sudakov effects and
the small-$x$ logarithms, and the associated evolution equations can be derived
as well. After resummation, we obtain the all order result as
\begin{eqnarray}
xG^{(DP)}(x,r_\perp,\zeta_c=Q)&=&-\frac{2}{\alpha_S}
\int \frac{d^2x_\perp d^2y_\perp}{(2\pi)^{4} }e^{ik_\perp\cdot r_\perp}{\cal H}^{DP}(\alpha_s(Q))e^{-{\cal S}_{pert}^g(Q^2,r_\perp^2)}\nonumber\\
&&\times {\vec{\bigtriangledown}}^2_{r_\perp}{\cal F}^{DP}_{Y=\ln 1/x}(x_\perp,y_\perp)\
,\label{resumdp}
\end{eqnarray}
where  ${\cal F}^{DP}_{Y=\ln 1/x}$ is renormalized dipole gluon distribution of Eq.~(\ref{fdp})
and ${\cal H}^{DP}=1+{\cal O}(\alpha_s)$.
The Sudakov factor follows the same as that for $WW$-gluon distribution in the last subsection,
where again $B_g^{(1)}=0$ by the same reason.

\section{Conclusion}

In summary, we have shown that one can treat both the Collins-Soper evolution and the BK evolution of  small $x$
parton TMDs simultaneously in a unified and consistent framework. To achieve so, we compute small
$x$ gluon TMDs in the Collins-11 scheme using the CGC approach. The resulting hard parts contain
two different type large logarithms which can be resummed by means of the Collins-Soper equation
and the BK equation, respectively. We emphasize that the relative size of two type large logarithms
is solely  determined by the kinematics of a physical process which involve three well separated
scales. The evolved small $x$ gluon TMD eventually can be expressed as the convolution of the
Sudakov form factor and the renormalized dipole amplitudes. In other words, at small $x$, gluon
TMDs can be matched onto dipole scattering amplitudes rather than the normal PDFs in the collinear
approach.

Our analysis is applied to the quark distribution, as well as the WW gluon TMD and the dipole gluon TMD cases. For the quark distribution and the WW gluon 
cases, we compute the parton TMDs involving a past pointing gauge link as well as these involving a
future pointing gauge link. Though it appears to be rather nontrivial, the DIS type and the
Drell-Yan type parton TMDs are shown to be identical as expected from the universality argument. As
a final remark, we also point out that in order to recover the correct Sudakov factor, it is
critical to calculate gluon TMDs using their matrix elements definition given in Eqs.(\ref{g1},
\ref{g2}) instead of these in Eqs.(\ref{e3}, \ref{e5}), 
since the gluon distributions defined in latter equations have already put gauge links on the light-cone.

Finally, we consider the present work as one more attempt to advance the study of the topical
issue: the interplay of small $x$ physics and TMD/spin physics. It shall provide a theoretical
ground for performing phenomenological analysis of the relevant physical observables that can be
measured at RHIC, LHC and the planned EIC.

{\bf Acknowledgments:}  This material is based upon work supported by the U.S. Department of Energy, 
Office of Science, Office of Nuclear Physics, under contract number 
DE-AC02-05CH11231, and within the framework of the TMD Topical Collaboration. 
This work is also supported by the National Science Foundations of
China under Grant No. 11675093, No. 11575070, and by the Thousand Talents Plan for Young Scientist.

\end{document}